\documentclass[10pt]{article}
\usepackage{amsmath}
\usepackage{float}
\usepackage[]{graphicx}
\usepackage{color}
\usepackage{marginnote}
\usepackage[UKenglish]{babel}
\usepackage{mathbbol}
\usepackage{textcomp}

\newcommand{\bis}{\textit{s}}
\newcommand{\bih}{\textit{h}}
\newcommand{\bn}{\textit{n}}
\newcommand{\bo}{\textit{o}}
\newcommand{\opt}{{\rm{opt}}}
\newcommand{\rx}{{\rm{x}}}

\begin{document}

\title{Measurement of the miscut angle in the determination of the Si lattice parameter}

\author{ C P Sasso, G Mana, and E Massa\\
INRIM -- Istituto Nazionale di Ricerca Metrologica\\
Strada delle Cacce 91, 10135 Torino, Italy\\
e.massa@inrim.it}
\maketitle
\begin{abstract}
The measurement of the angle between the interferometer front mirror and the diffracting planes is a critical aspect of the Si lattice-parameter measurement by combined x-ray and optical interferometry. In addition to being measured off-line by x-ray diffraction, it was checked on-line by transversely moving the analyser crystal and observing the phase shift of the interference fringe. We describe the measurement procedure and give the miscut angle of the $^{28}$Si crystal whose lattice parameter was an essential input-datum for, yesterday, the determination of the Avogadro constant and, today, the kilogram realisation by counting atoms. These data are a kindness to others that might wish to repeat the measurement of the lattice-parameter of this unique crystal.
\end{abstract}


%\ioptwocol

\section{Introduction}
The measurement of the surface orientation was a critical aspect of the $^{28}$Si lattice-parameter measurement by combined x-ray and optical interferometry \cite{Basile_1991,Massa_2011,Massa_2015}. It refers to the parallelism of the diffracting planes to the mirror-polished (front and rear) surfaces of the analyser crystal, whose displacement is measured in terms of a calibrated optical wavelength.

To achieve a fractional accuracy near to nine significant digits in the measurement of the lattice parameter, we polished the (front and rear) analyser surfaces parallel to the $\{220\}$ diffracting planes to within a few microradians. To this end, we realised an apparatus whereby the $^{28}$Si monocrystal was ground and optically polished and the residual misalignment was measured \cite{Bergamin_1999}. Subsequently, we checked the miscut angle on-line directly by the combined x-ray and optical interferometer.

The paper is organised as follows. After a con\-ci\-se description of combined x-ray and optical interferometry, in sections \ref{errors} and \ref{alignment}, we model the pro\-ject\-ion errors and detail how the x-ray in\-ter\-ferometer is aligned to compensate them. Next, sections \ref{diffraction} and \ref{x-ray} describe the measurement procedures of the miscut angle. The results are given in section \ref{results}.

\section{X-ray interferometry}
\begin{figure}\centering
\includegraphics[width=0.99\columnwidth]{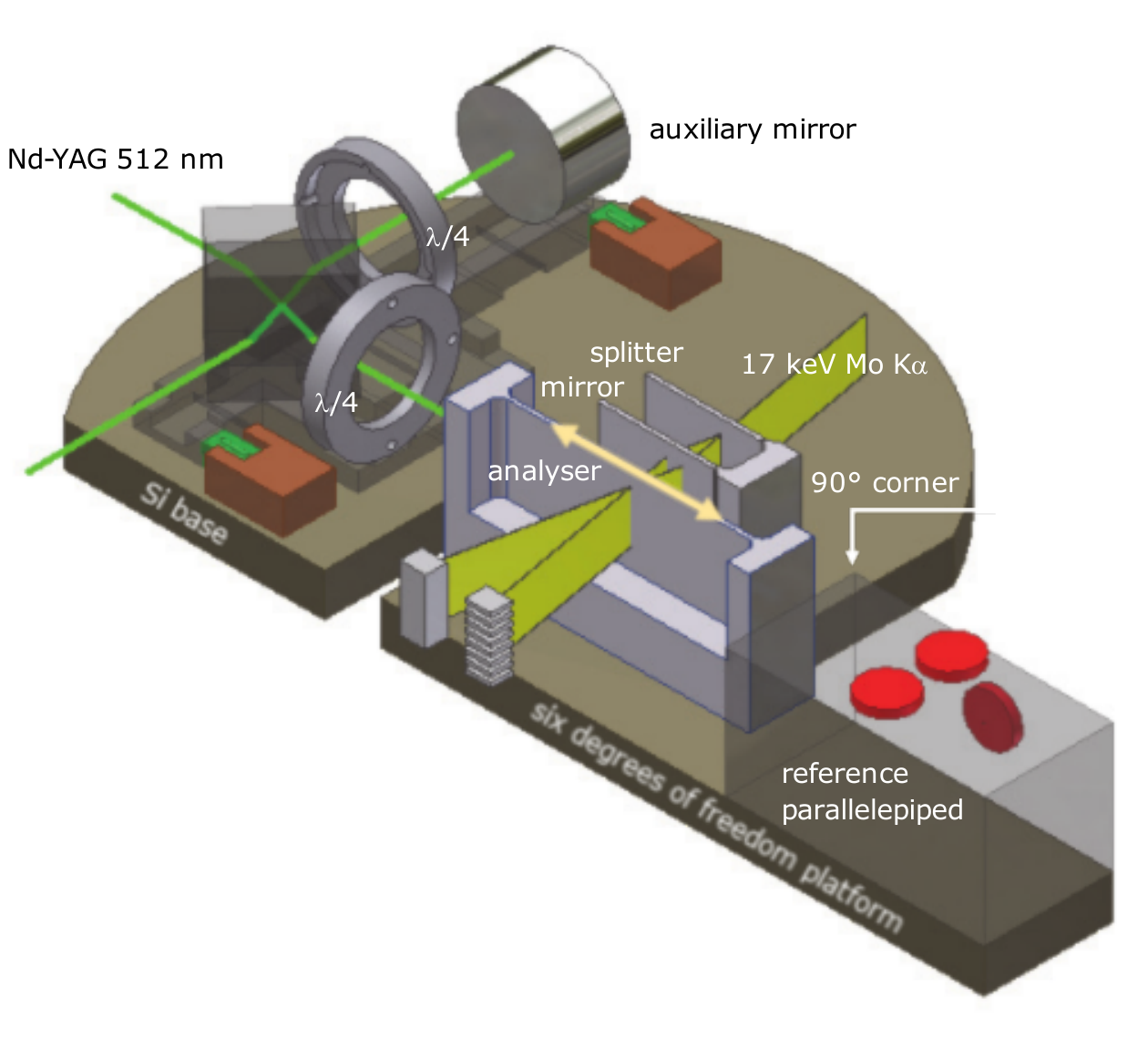}
\caption{INRIM's combined x-ray and optical in\-ter\-ferometer. The analyser displacement (yellow arrow) is measured {\it vs.} an auxiliary mirror mounted on the same Si base as the fixed splitter-mirror pair of the x-ray interferometer. To achieve pico\-metre resolution, the optical interferometer uses polarisation en\-coding and phase modulation. The pitch and yaw angles are mea\-sured to within nanoradian resolution via differential wavefront sensing. The transverse displacements (horizontal and vertical) and roll angle are measured via a (coated) glass parallelepiped (having a 90$^\circ$ corner) and capacitive sensors (red discs), which are integral with the fixed Si base. Feedback loops set the movement orthogonal to the parallelepiped front.} \label{fig01}
\end{figure}

Our x-ray interferometer is like a Mach-Zehnder in\-ter\-ferometer of visible optics \cite{Vittone:1997a}. As shown in Fig.\ \ref{fig01}, it splits and recombines 17 keV x rays by Laue diffractions in perfect Si crystals. The splitter, mirror, and analyser operate in symmetric Laue geometry, where the $\{220\}$ diffracting planes are perpendicular to their surfaces. When the interfering beams are phase-shifted by moving the analyser orthogonally to the diffracting planes, interference fringes are observed, their period being the plane spacing, $d_{220} \approx 192$ pm.

The analyser displacement is measured by a laser interferometer having picometre sensitivity and accuracy. The lattice spacing is measured as $d_{220} = m\lambda/(2n)$, where $n$ is the number of x-ray fringes observed in $m$ optical orders of $\lambda/2$ period. Since the Si thermal expansion is about $2.5\times 10^{-6}$ 1/K, the temperatures of the two interferometer crystals must be equal to within a few millikelvin. Therefore, to ensure temperature uniformity and stability and to eliminate the adverse influence of the refractive index of air, the apparatus is hosted in a (passive) thermovacuum chamber.

The analyser is displaced using a guide where an L-shaped carriage slides on a quasi-optical rail. An active platform (see Fig.\ \ref{fig01}) with three piezoelectric legs rests on the carriage. Each leg expands vertically and shears in the two transverse directions, thus allowing compensation for the sliding errors over six degrees of freedom to atomic-scale accuracy. The analyser displacement, parasitic (pitch, yaw, and roll) rotations, and transverse (horizontal and vertical) motions are measured via laser interferometry, differential wavefront sensing, and capacitive transducers. Eventually, feedback loops provide picometer positioning, nanoradian alignment, and movements with nanometer straightness.

\subsection{Projection errors}\label{errors}
Since, the x-ray and optical interferometers project the analyser displacement, $\bis$, into the unit normals to the diffracting planes and front mirror,  $\bih$ and $\bn$, to ensure that both projections are equal, the analyser front mirror was polished as parallel as possible to the diffracting planes. Also, the displacement was electronically servoed to bisect the misalignment angle, and the laser beam was aligned as much as possible orthogonally to the front mirror.

\begin{figure}\centering
\includegraphics[width=0.9\columnwidth]{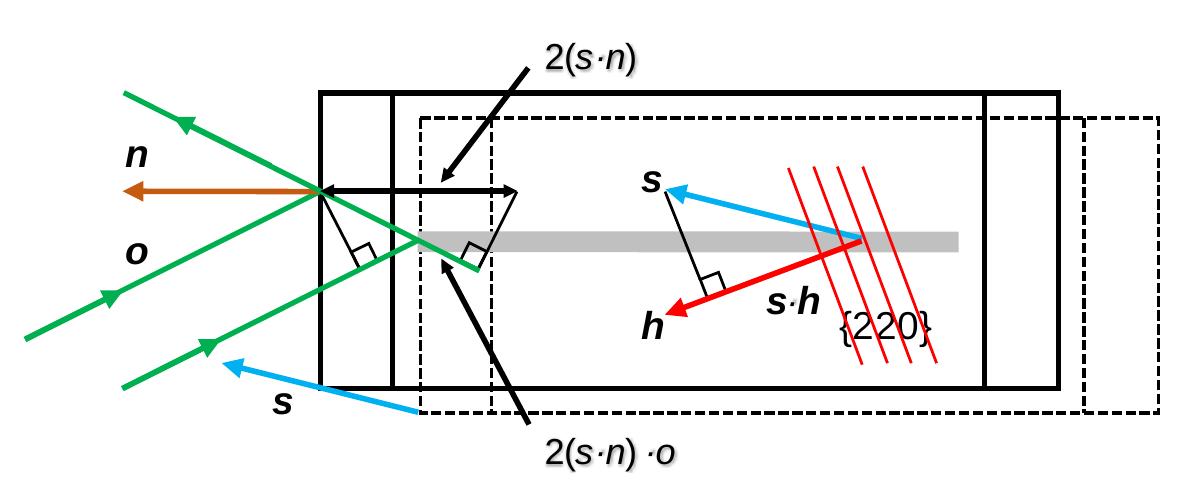}
\caption{Two-dimensional sketch of a hypothetical arrangement of the vectors determining the projection errors. The dashed frame shows the analyser before the displacement $\bis$. Green: laser beam, brown: normal to the analyser front mirror, blue: displacement (not in scale), red: normal to the lattice planes.} \label{fig02}
\end{figure}

A model of the projection errors is as follows. In a plane-wave approximation, as shown in Fig. \ \ref{fig02}, the mea\-sure\-ment equation of the laser interferometer,
\begin{equation}\label{OINT}
 s_\opt = 2(\bis \cdot \bn)(\bn \cdot \bo) ,
\end{equation}
where $\bo$ is a unit vector parallel to the beam axis, is two times the component of the analyser displacement along the front-mirror normal, $\bis \cdot \bn$, projected on the beam axis.

\begin{figure}\centering
\includegraphics[width=0.8\columnwidth]{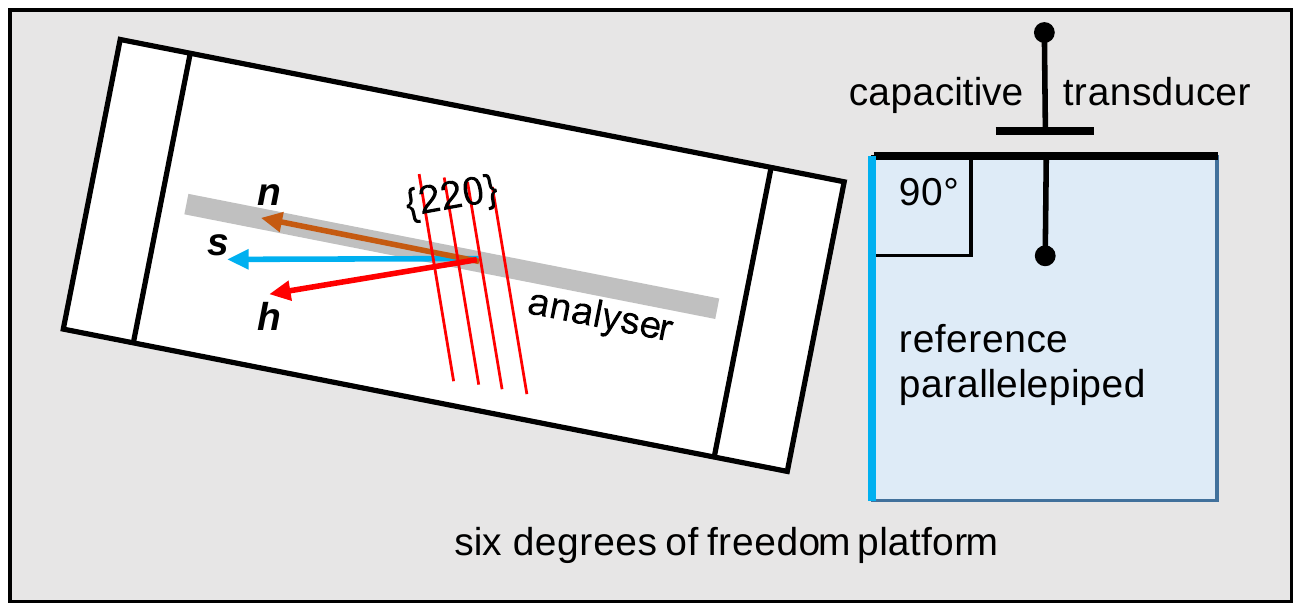}
\caption{Two-dimensional sketch of a hypothetical alignment of the analyser crystal. A feedback loop locks the reading of the capacitive transducer; therefore, the analyser displacement occurs orthogonally to the front (sky-blue) face. Brown: normal to the analyser front mirror, blue: displacement, red: normal to the lattice planes.} \label{fig03}
\end{figure}

The measurement equation of the x-ray interferometer is $s_\rx = \bis \cdot \bih$. Therefore, considering separately the projections on the laser beam, $\bn \cdot \bo$, and on the front-mirror normal, $\bis \cdot \bn$, the projection errors compensate if $\bis \cdot \bn = \bis \cdot \bih$. If $\bn=\bih$, the displacement direction is irrelevant.  Otherwise, it is required that
\begin{equation}
 \bis = s_\| \widehat{(\bn + \bih)} + s_\bot \widehat{(\bn \times \bih)} ,
\end{equation}
where the hats indicate normalised vectors and the $s^2 = s_\|^2 + s_\bot^2$ decomposition is arbitrary. Otherwise said, $\bis$ must lie in the plane bisecting the normals to the diffracting planes and front mirror. The fractional measurement-error, $\hat{\bis}\cdot(\bn-\bih)$, is proportional to the miscut angle, $|\bn-\bih|$, and bisection error, $\hat{\bis}\cdot\widehat{(\bn - \bih)}$.

\subsection{Alignment procedure}\label{alignment}
To achieve a nine-digit accuracy, the miscut angle $|\bn-\bih|$ must be as small as possible.  At the same time, the normals to the front mirror and laser beam, $\bn$ and $\bo$, and the movement direction, $\hat{\bis}$, must be aligned each other so that $\hat{\bis}\cdot\widehat{(\bn - \bih)}$ is minimum.

Figure \ref{fig03} shows the way we aligned the x-ray interferometer. A glass parallelepiped rests on the same platform as the analyser (see Fig.\ \ref{fig01}). It has the top and right-side optically poli\-shed to within $\lambda/20$ flatness and metallic coated. Capacitive sensors (rendered as red discs in Fig.\ \ref{fig01}) monitor its horizontal and vertical positions and drive piezoelectric actuators to keep them fixed to within sub-nanometer accuracy. These feedback loops ensure that the parallelepiped moves orthogonally to its front, which is orthogonal to the top and side to within 12 $\mu$rad and 8 $\mu$rad, respectively.

As shown in Fig.\ \ref{fig03}, the analyser is mounted in such a way the $\{220\}$ planes and front mirror are symmetrically placed to the parallelepiped front. This alignment is made with the aid of an autocollimator looking at the parallelepiped front and the analyser rear. Obviously, the analyser miscut and parallelepiped orthogonality are taken into account.

\begin{figure}\centering
\includegraphics[width=0.9\columnwidth]{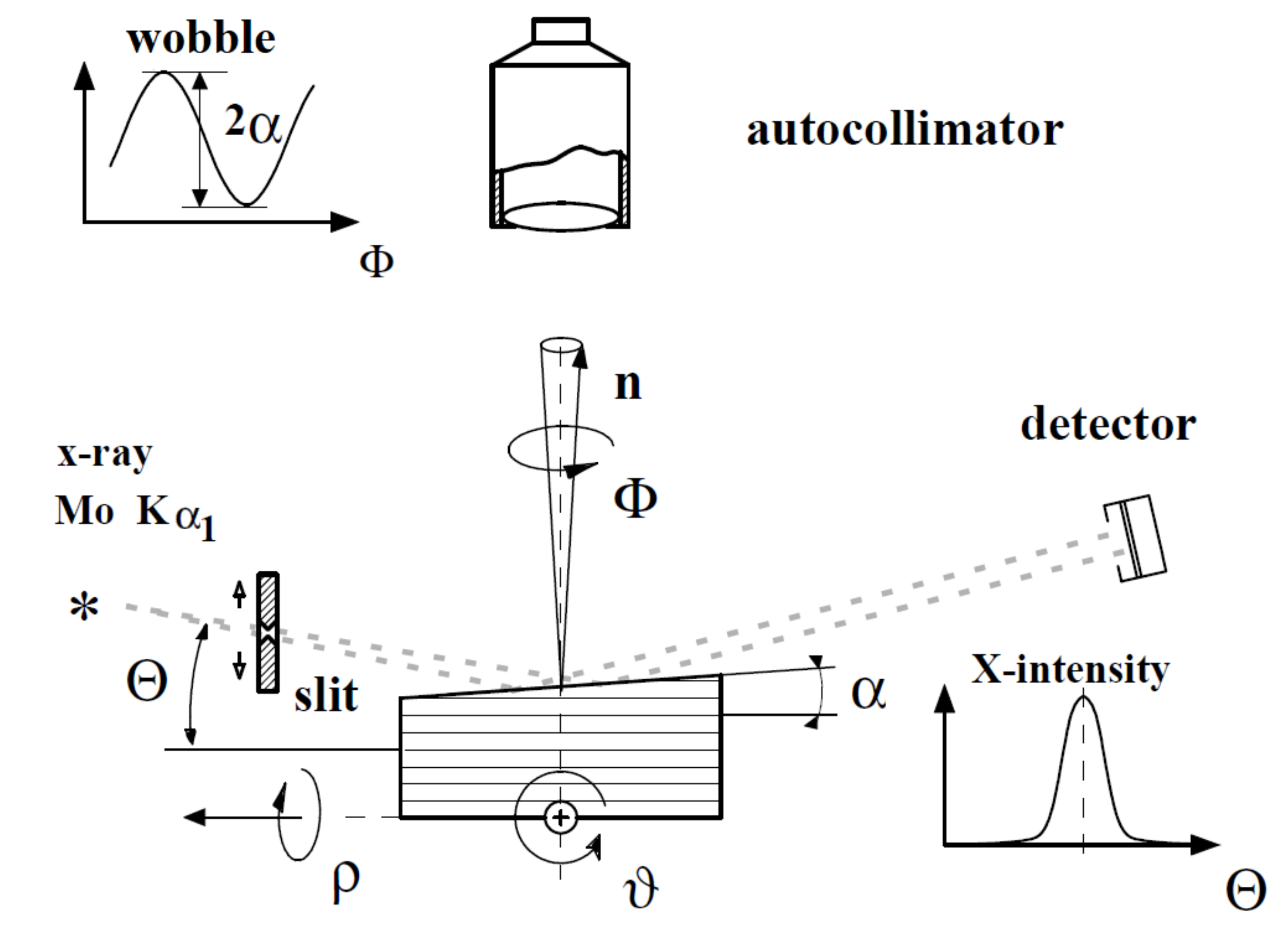}
\caption{Diagram of the apparatus for crystal-orientation and measurement of the miscut angle (adapted from \cite{Bergamin_1999}).} \label{fig04}
\end{figure}

\begin{figure}\centering
\includegraphics[width=0.9\columnwidth]{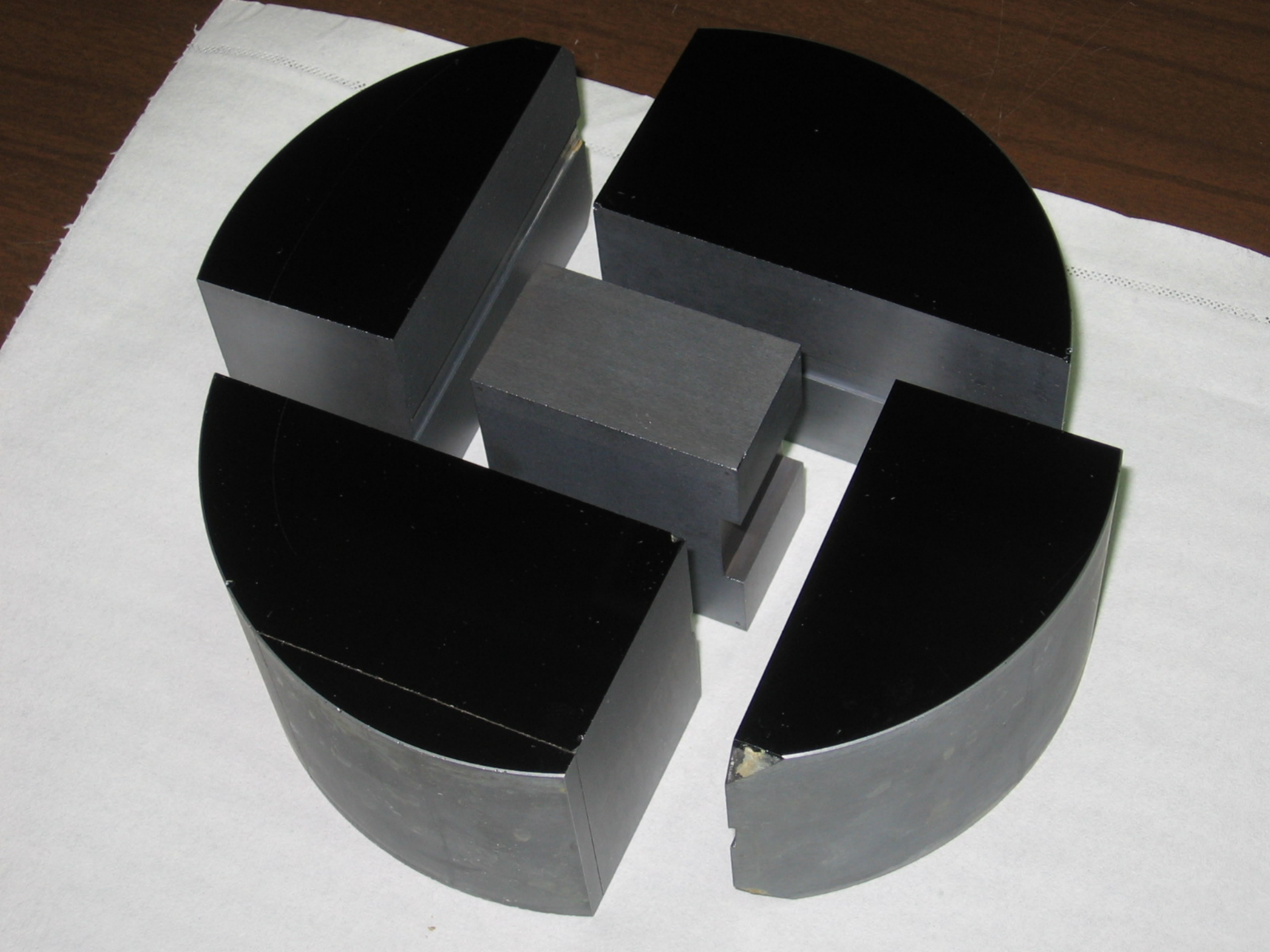}
\includegraphics[width=0.9\columnwidth]{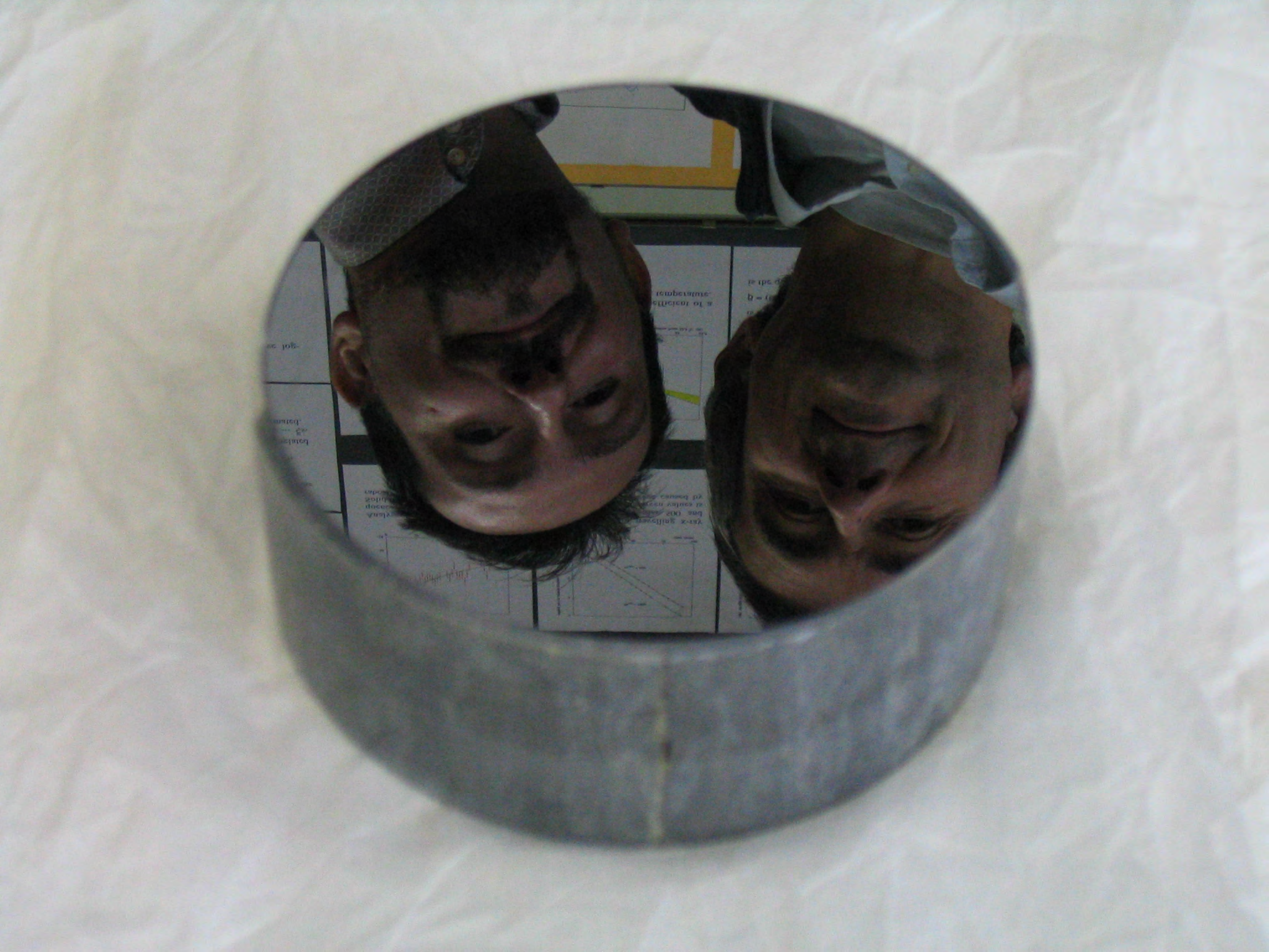}
\caption{Photograph of the silicon disc embedding the $^{28}$Si block for the optical polishing of the front and rear surfaces. From this $^{28}$Si block, both the INRIM's and PTB's interferometers were manufactured.} \label{fig512}
\end{figure}

\section{Measurement of the miscut angle}
\subsection{X-ray diffraction}\label{diffraction}
The raw crystal -- a $^{28}$Si block $(55 \times 60 \times 28)$ mm$^3$ -- was oriented by x-ray diffraction. Two opposite faces -- which will be the interferometer front and rear mirrors -- were ground and optically polished parallel to the $\{220\}$ lattice planes.

Figure \ref{fig04} shows how the angles between the lattice planes and the crystal surfaces were measured. The main components of the apparatus are a crystal-holder made by a two-axis tilter integrated into a precision rotary table and an autocollimator measuring the wobble of the crystal surface. Figure \ref{fig512} shows the silicon disc embedding the $^{28}$Si block (having the same crystallographic orientation) for the orientation and optical polishing.

By tilting the crystal to keep aligned the Bragg's reflection of the x rays when it is rotated, the lattice planes are set orthogonal to the rotation axis. Next, the autocollimator measures the wobble of the optically-polished surface. More details are given in \cite{Bergamin_1999}. The accuracy achieved is near to 1 $\mu$rad, depending on the surface flatness: a 2 cm wide part of a $\lambda/20$ surface has a sagitta of 5 nm and an end-to-end tilt of 5 $\mu$rad.

\begin{figure}\centering
\includegraphics[width=0.8\columnwidth]{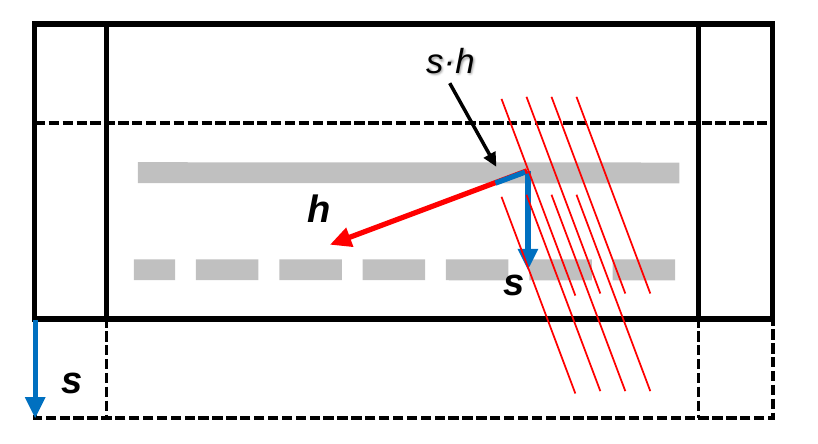}
\caption{Schematic diagram of the miscut-angle measurement (horizontal component) by combined x-ray and optical interferometry. The dashed frame shows the analyser after a transverse (horizontal)  displacement $\bis$. A feedback loop uses the optical-interferometer signal to lock to zero the axial displacement and ensures that $\bis$ lies in the mirror surface. The phase shift of the x-ray interference before and after the displacement is $\bis \cdot \bih$.} \label{fig06}
\end{figure}

\subsection{X-ray interferometry}\label{x-ray}
We repeated the measurement of the miscut angle on-line directly by the combined x-ray and optical interferometer. This measurement repetition strengthened our confidence in the measured angle and checked the miscut of the surface area actually sensed by the laser beam.

As shown in Fig.\ \ref{fig06}, the measurement was done by translating the analyser transversally, in the vertical and horizontal directions. Because of the present limited operation range of the supporting platform, the translations were limited to a few micrometres. To be safe, we verified not to lose integer orders of the x-ray interference by carrying out preliminary displacements of increasing amplitude.  To ensure that the translations occur in the plane of the front mirror, a feedback loop locks to zero (to within 1 pm and 1 nrad) the axial displacement and the pitch and yaw rotations.

\begin{figure}\centering
\includegraphics[width=0.99\columnwidth]{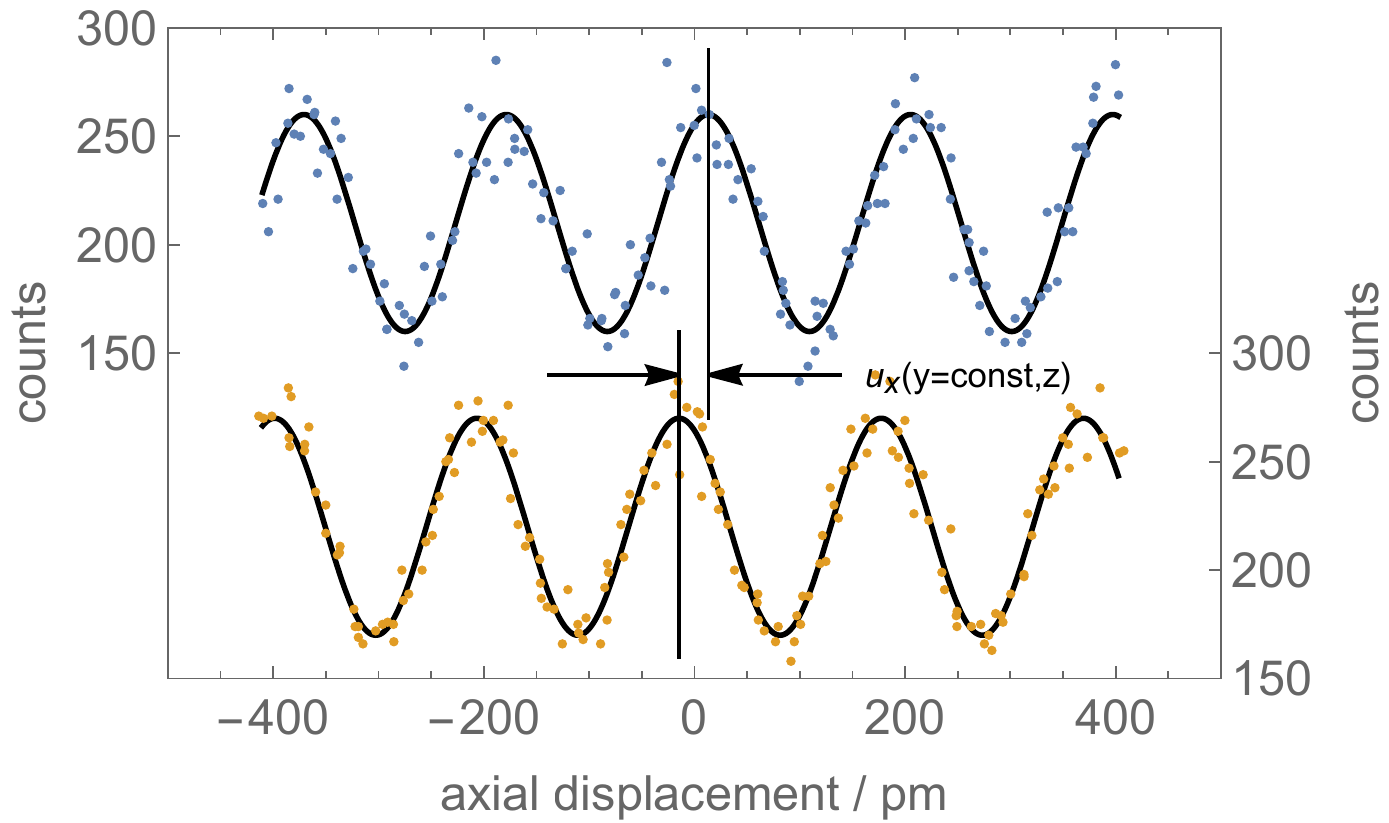}
\caption{Scans of the x-ray fringes before (orange) and after (blue) a vertical analyser displacement $s_y = 3.2(3)$ $\mu$m of the analyser. The dots are the x-photons counted in 100 ms. The solid lines are the best-fit sinusoids approximating the data. The observed phase difference is $u_x(y={\rm const.},z) = 0.145(18)\, d_{220}$ (see the first two measured phases in Fig.\ \ref{fig07}).} \label{fig07}
\end{figure}

\begin{figure}\centering
\includegraphics[width=0.99\columnwidth]{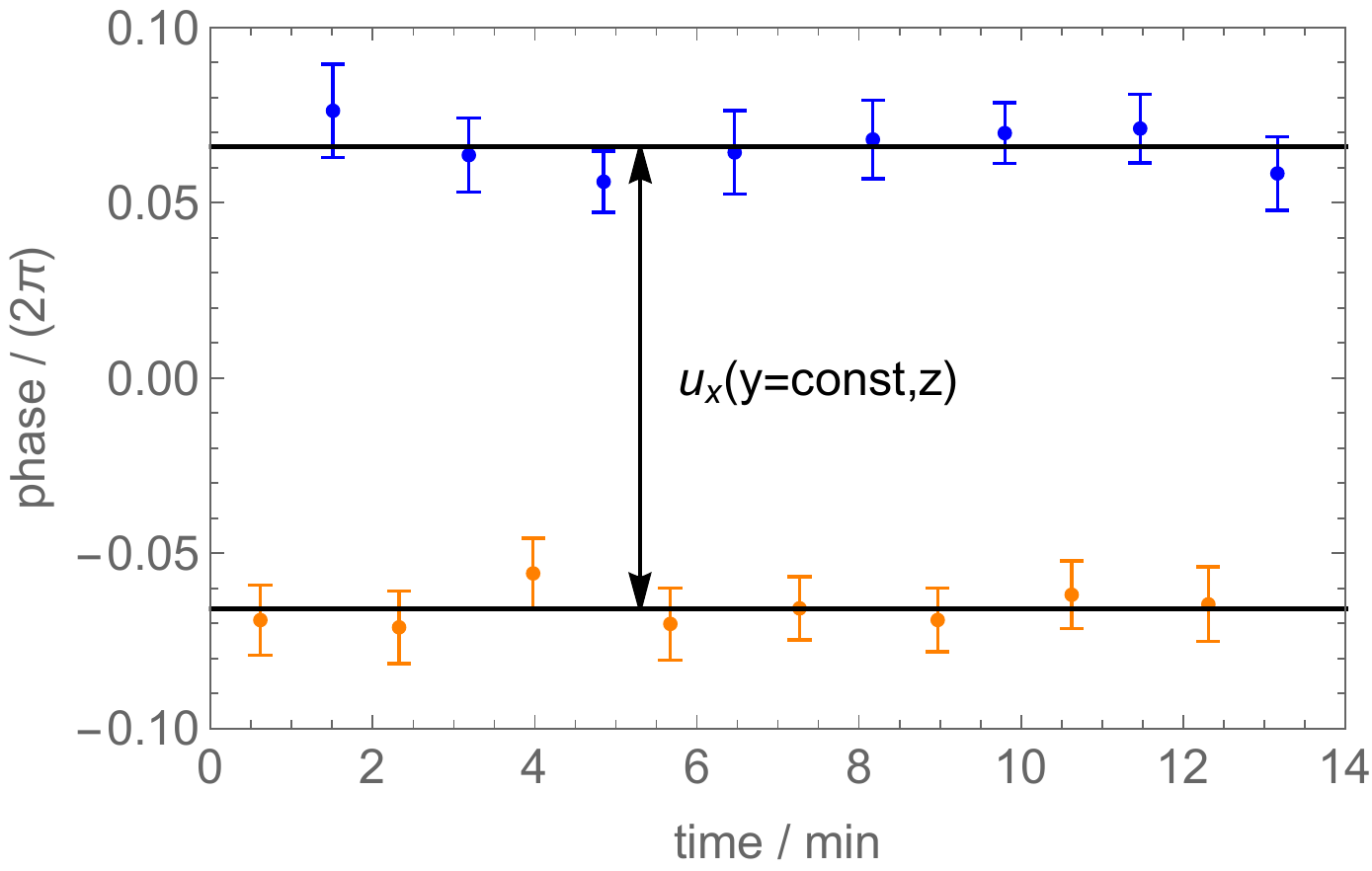}
\caption{Phases of the x-ray fringes measured before (orange) and after (blue) a vertical analyser displacement of 3.2(3) $\mu$m. The dots are the measured phases (see Fig.\ \ref{fig07}). The bars are the associated standard uncertainties. The same best-fit polynomial has been subtracted from the measured phases. The observed difference is $u_x(y={\rm const.},z) = 0.135(6)\, d_{220}$, to which a miscut angle $\epsilon_{xy} = 8.1(8)$ $\mu$rad will correspond.} \label{fig08}
\end{figure}

Figure \ref{fig07} shows the x-ray fringes observed before and after a vertical displacement of the analyser equal to 3.2 $\mu$m. As shown in Fig.\ \ref{fig06}, the phase difference between the x-ray fringes detected at the start and end positions are proportional to the lattice displacements at the end, $u_x(y,z=\rm{const.})=\bis_y \cdot\bih$ and $u_x({y=\rm{const.}},z)=\bis_z \cdot\bih$, relative to the start.

Since it was not possible to eliminate the drift between the optical and x-ray signals, the analyzer was moved repeatedly forwards and backwards, and the two signals were repeatedly sampled at each end. Next, the drift was identified and subtracted by fitting the phases of the x-ray fringes with polynomials differing only by the sought phase difference. Figure \ref{fig08} shows the results.

After calculating the ratios to the displacements, we obtain the shear strains $\epsilon_{xy} = u_x(y,z=\rm{const.})/s_y$ and $\epsilon_{xz} = u_x({y=\rm{const.}},z)/s_z$, which are nothing else that the horizontal and vertical components of the sought miscut angle.

\section{Results}\label{results}
\begin{figure}[H]\centering
\includegraphics[width=0.8\columnwidth]{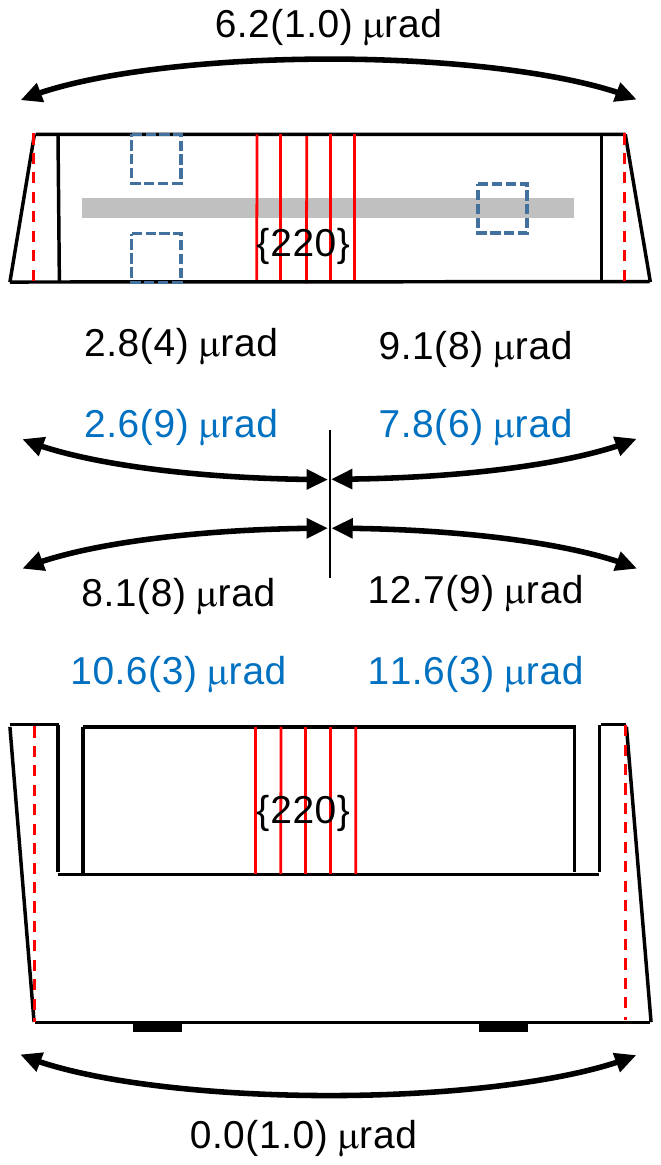}
\caption{Measured values of the miscut angles of the INRIM's $^{28}$Si interferometer; the analyser sketch shows the angle signs. Blue: x-ray diffraction. Black: combined x-ray and optical interferometry. The lattice planes are assumed everywhere parallel better than 0.1 $\mu$rad. The measured components of the angle between the front and rear mirrors are also given.} \label{fig09}
\end{figure}

The measurement results are given in Fig.\ \ref{fig09}. The horizontal and vertical components of the angle between the front and rear mirrors were measured with the aid of a precision rotary table having a sub-arcsecond resolution, a calibrated polygon, and an autocollimator.

The measured values of the front-to-rear angles are systematically smaller than the values inferred by the diffractometric and interferometric measurements, which are in good agreement. We did not in\-ves\-ti\-gate the origin of this discrepancy; besides, it was irrelevant in aligning the x-ray and optical interferometers and correcting the lattice parameter measurement. However, together with the potentially higher resolution achievable, it might be a clue of the superior accuracy of Si-monocrystal polygons having the angles calibrated to the lattice planes \cite{Becker_1990}.

\section{Conclusions}

The measurement of the $^{28}$Si lattice parameter in terms of optical wavelengths opened the way to determine the Avogadro and Planck constants and to realise the kilogram by counting $^{28}$Si atoms.

We reported about two independent measurements of the miscut angle of the front a rear surface of the $^{28}$Si block from which the x-ray interferometer used for this measurement was obtained. A second x-ray interferometer has been recently manufactured from the same block, and an additional lattice parameter measurement is underway \cite{Birk_2020}.

The value of this angle is critical to quantify the projection errors of the x-ray and optical interferometers and to align the crystal under measurement in such a way to make these projections identical. Therefore, it lets these x-ray interferometers be future proof, in that who may wish to remeasure the lattice spacing of these unique crystals does not need to redetermine the miscut angle.

%\section*{References}
\bibliography{miscut}

\end{document}